\documentclass[useAMS,usenatbib]{mn2e}

\usepackage[ansinew]{inputenc}
\usepackage{graphicx} 
\usepackage{wasysym}

\def \aj {AJ}
\def \mnras {MNRAS}
\def \apj {ApJ}
\def \apjs {ApJS}

\def \aap {A\&A}
\def \aaps {A\&AS}

\def \pasp {PASP}

\def \angstron {$\rm \AA$}
\def \Teff {T$_{\rm eff}$}

\title[Testing the Accuracy of Synthetic Stellar Libraries]{Testing the 
Accuracy of Synthetic Stellar Libraries}

\author[Martins \& Coelho]{Lucimara P. Martins$^{1}$\thanks{E-mail:
lucimara@astro.iag.usp.br (LM), pcoelho@usp.br (PC).}, Paula Coelho$^{1,2}$\\
$^1$ Universidade de S\~ao Paulo, IAG, Rua do Mat\~ao 1226,
 S\~ao Paulo 05508-900, Brazil;\\
$^2$ Max-Planck-Institut f\"ur Astrophysik, Karl-Schwarzschild-Strasse 1, 85740 
Garching bei M\"unchen, Germany \\
}

\begin{document}

\date{}

\pagerange{\pageref{firstpage}--\pageref{lastpage}} \pubyear{2007}

\maketitle

\label{firstpage}

\begin{abstract}

One of the main ingredients of stellar population synthesis
models is a library 
of stellar spectra. Both empirical and theoretical libraries are used for this 
purpose,  and the question about
which one is preferable is still debated in the literature. Empirical and 
theoretical
libraries are being improved significantly over the years, and many
libraries have become available lately. However, it is not clear in the 
literature
what are the advantages of using each of these new libraries, and how far
behind are models compared to observations. Here we compare in detail some of 
the
major theoretical libraries available in the literature with observations,
aiming at detecting weaknesses and strengths from the stellar population 
modelling point of view. Our test is twofold: we compared model predictions
and observations for broad-band colours and for high resolution
spectral features. Concerning the broad-band colours, we measured
the stellar colour given by three recent sets of model atmospheres and 
flux distributions \citep{ATLASODFNEW,MARCS05,PHOENIX05}, 
and compared them with a recent
UBVRIJHK calibration \citep{worthey_lee07} which is mostly based on empirical data.
We found that the models can
reproduce with reasonable accuracy the stellar colours for a fair interval
in effective temperatures and gravities. The exceptions are: 1) the U-B colour,
where the models are typically redder than the observations, and; 2) the
very cool stars in general (V-K $\apprge$ 3).  
\citet{ATLASODFNEW} is the set of models that best reproduce the
bluest colours (U-B, B-V) while \citet{MARCS05} and \citet{PHOENIX05}
more accurately predict the visual colours. The three sets of models perform 
in a similar way for the infrared colours. Concerning the high resolution
spectral features, we measured 35 spectral indices defined in the literature
on three high resolution
synthetic libraries \citep{coelho+05,martins+05,munari+05}, and compared
them with the observed measurements given by three empirical libraries
\citep{INDOUS,MILES,ELODIE}. The measured indices 
cover the wavelength range from 
$\sim$ 3500 to $\sim$ 8700 {\angstron}.
We found that the direct comparison
between models and observations is not a simple task, given the
uncertainties in parameter determinations of empirical libraries.
Taking that aside, we found that in general the three libraries present
similar behaviours and systematic deviations. 
For stars with {\Teff} $\leq$ 7000K,
the library by \citet{coelho+05} is the one with best 
average performance.
We detect that lists of atomic and molecular
line opacities still need improvement, specially in the blue region of the
spectrum, and
for the cool stars ({\Teff} $\apprle$ 4500K). 
\end{abstract}


\section{Introduction}
Evolutionary population synthesis models describe the 
spectral evolution of stellar systems, and are fundamental 
tools in the analysis of both nearby and distant 
galaxies 
\citep[e.g.][]{BC03,cervino_hesse94,fioc_rocca97,leitherer+99,vazdekis99,
buzzoni02,jimenez+04,pegase-hr,delgado+05,maraston05,schiavon06}.
They are needed to derive the star formation history
and chemical enrichment in a 
variety of systems, from early 
type galaxies and spiral bulges to star forming galaxies at different 
redshifts.
 
Libraries of stellar spectra are one of the main ingredients of stellar 
population models, 
and both empirical and theoretical 
libraries have improved dramatically in recent years, allowing
the construction of more detailed models. 
Observations are also becoming increasingly better and 
demanding more from the modelling point of view. 

Recently, many new empirical libraries
suitable to stellar population synthesis have been made 
available with improved spectral
resolution and parameter coverage: e.g. 
STELIB \citep{stelib},
UVES POP \citep{uvespop},
Indo-US \citep{INDOUS},
Elodie \citep{ELODIE},
MILES \citep{MILES},
and NGSL \citep[][]{gregg+04}.

The choice of using either an empirical or a synthetic library in stellar 
population models is a subject of debate. 
Many aspects are important when considering a library for stellar population
synthesis, and parameter coverage is one of the main issues. A good parameter coverage 
is not trivial for empirical libraries, which are limited to whatever is
possible to obtain given observational constraints (resolution, wavelength coverage,
exposure time, etc.).
They have to cover not only a good wavelength
range (which limits the spectral resolution), but also cover from cool to hot 
stars, dwarfs and giants, and different chemical abundances. 

Amongst the synthetic libraries, perhaps the most widely used is the 
flux distribution predicted by the
\citet{kurucz93}
model atmospheres. 
The BaSeL library \citep{basel1,basel2,basel3} extended these flux distributions 
including spectra of M stars computed with model atmospheres 
by
\citet{fluks+94}, \citet{bessel+89,bessel+91} and \citet{allard_hauschildt95}.
However the spectral 
resolution of the 
BaSeL library is limited to $\sim$ 
20~{\angstron}, 
which is by far lower than the modern observed spectra of both individual 
stars and integrated stellar populations.
Resolution ceased to be a limitation recently, with many 
high-resolution theoretical libraries appearing in the literature 
\citep{chavez+97,barbuy+03,bertone+03,lanz_hubeny03,
zwitter+04,murphy_meiksin04,
coelho+05,MARCS05,PHOENIX05,uvblue05,malagnini+05,martins+05,munari+05,
nemo06}.
Many
of these libraries were created with refined and updated line lists,
state of the art model atmospheres and spectral synthesis codes, and
a very extensive parameter coverage. A qualitative comparison of some of the recent high
resolution synthetic libraries is given by \citet{bertone05}.

The major concern when using 
synthetic libraries for high 
resolution stellar population models is to know whether a synthetic 
library can safely 
replace an empirical one. 
These libraries are based on model atmospheres and therefore are
limited to the approximations adopted in the computations.
Ideally, one would like to generate models
that accounts for all the effects taking place across the HR diagram:
non-Local Thermodynamic Equilibrium (NLTE), line-blanketing, sphericity,
expansion, non-radiative heating,
convection, etc. Such an approach is unfeasible at present time, even if
the astrophysical models were available. What is usually done is
to take some of these effects into account where they matter the most.
The hardest stars to reproduce in this sense are the 
very hot and very cool stars, where
extreme scenarios take place (e.g. non-LTE effects for very hot stars, 
and sphericity for cool giants).
Additionally, computing reliable high-resolution 
synthetic spectra is a very challenging task, since it requires building an 
extensive and accurate list of atomic and molecular line opacities. 

Nevertheless, synthetic libraries overcome limitations of empirical libraries, for
instance their inability to cover the whole space in atmospheric parameters, 
and in particular abundance patterns that differ 
from that of the observed stars (mainly from the solar neighbourhood, and in 
some few cases from the Magellanic Clouds). Therefore, population models based 
solely on empirical libraries cannot reproduce the 
integrated spectra of systems that have undergone star formation histories 
different than the Solar Neighbourhood.

With so many different choices for the stellar library, 
the stellar population modeller might
feel lost about which library should be used. It is certain that each
of these libraries have its own strengths and weaknesses, but identifying
them is not always trivial. We propose in this work to make a
detailed comparison between some of the major synthetic stellar libraries 
available,
comparing them against empirical libraries. 

This paper is organised as follows:
in \S 2 we present an overview of theoretical libraries.
In \S 3 the model predictions of three sets of model atmospheres 
\citep{ATLASODFNEW,MARCS05,PHOENIX05}
for broad-band colours are compared to the
empirical UBVRIJHK relation from \citet{worthey_lee07}.
In \S 4 we compare model spectral indices predicted by three recent
high-resolution libraries \citep{coelho+05,martins+05,munari+05} 
to indices measured in the empirical libraries by \citet{INDOUS,MILES,ELODIE}.

For the
purpose of the present work, we focus our comparisons on the solar metallicity
regime, where the completeness of the empirical libraries is higher, as well as 
the accuracy of the stellar atmospheric parameters.
Our conclusions and discussions are presented in \S 5.

\section{Overview of the Theoretical Libraries}

The nomenclature used by atmosphere and synthetic spectra modellers are
sometimes confusing for the stellar population models users.

By {\it model atmosphere} we mean the run
of temperature, gas, electron and radiation pressure, 
convective
velocity and flux, and more generally, of all relevant quantities
as a function of some depth variable (geometrical, or optical depth
at some special frequency, or column mass).
The {\it flux distribution} or {\it synthetic spectra} is the 
emergent flux predicted
by a model atmosphere, and is required for comparison
with observations.

It is convenient from the computational point of view to split the calculation
of a synthetic spectra in two major steps: the calculation of the model
atmosphere, commonly adopting Opacity Distribution Function technique
\citep[ODF,][]{strom_kurucz66} $-$
and the calculation of the emergent flux with a spectral synthesis code.

Alternatively, model atmosphere codes that use 
an Opacity Sampling (OS) method
to account for the line absorption \citep[e.g.][]{johnson_krupp76} 
can directly produce as output
a well sampled flux distribution. The OS technique is more time 
consuming from the computational point of view then the ODF technique, 
but allows for a 
much larger flexibility in modelling. For example, peculiar chemical
compositions  can be easily consideredthat'.

The majority of model atmospheres available are
1D and hydrostatic, assume LTE and treat convection 
with the mixing length theory. 
The mixing length theory was 
introduced in ATLAS6 code
by \citet{kurucz79}, and is a phenomenological approach to convection
in which it is assumed that the convective energy
is transported by eddy ``bubbles" of just one size.
t requires an adjustable parameter
$\alpha_{ML}$, which represents the ratio between the characteristic length 
(distance travelled by an element of fluid before its dissolution)
and the scale height of the local pressure  (H$_{p}$). The parameter
$\alpha_{ML}$ has to be set at different values to fit different types
of observations 
\citep{steffen_ludwig99},
and no single value
works well in all classes. An alternative convective model is 
Full Spectrum Turbulence, introduced by \citet{canuto_mazzitelli91} 
and adopted, for example, by NeMo grid of atmospheres \citep{heiter+02}.

Throughout this paper we further distinguish a {\it flux distribution} from 
a {\it synthetic spectrum}.
The {\it flux distribution} is the spectral energy distribution predicted directly 
by a model atmosphere, and is commonly available together with the model
atmospheres. This is
the case, for example, of the synthetic libraries by 
\citet[][]{ATLASODFNEW},
\citet[][]{PHOENIX05} and
\citet[][]{MARCS05}.

By {\it synthetic spectrum} we mean the flux calculated by a
line profile synthesis code, using as input a model atmosphere and 
a refined atomic and molecular line list, that can be at some 
extend different from the line list adopted in the model atmosphere computation.
It can also adopt different
chemical compositions than the model atmosphere in order to account for small
variations in the abundance pattern (as long as the difference is not enough
to produce important changes in the structure of the atmosphere). This is
the method commonly used in high resolution stellar spectroscopy studies,
and it is the case of the libraries from
\citet[][]{coelho+05},
\citet[][]{martins+05} and
\citet[][]{munari+05}.
A synthetic spectrum is usually computed at a higher resolution than a model
atmosphere flux distribution, given that it aims at resolving individual 
line profiles. 

Additionally, a theoretical library that is intended to produce accurate high resolution line 
profiles is not generally a library that also predicts 
good spectrophotometry. That happens because 
usually only the lower lying energy levels of atoms have been determined in
laboratory. If only those transitions are taken into account in a model
atmosphere, the line blanketing
would be severely incomplete. To avoid this deficiency and to improve
both the temperature structure of the model atmospheres and the spectrophotometric
flux distributions, the computation requires accounting for lines where one
or both energy levels have to be predicted from quantum mechanical
calculations. These so-called 
``predicted lines" \citep[hereafter PLs,][]{kurucz92} are an essential contribution
to the total line blanketing 
in model atmospheres and flux distribution computations. But as the 
theoretical predictions are accurate to
only a few percent, wavelengths and computed intensities for these 
lines may be largely uncertain. As a consequence the PLs
may not correspond in position and intensity to the observable 
counterparts \citep{bell+94,castelli+kurucz04}, ``polluting" the high 
resolution synthetic spectrum. Therefore, synthetic libraries
that are aimed at high resolution studies do not include the PLs, and
thus they provide less accurate spectrophotometric predictions when compared
to the 
flux distributions libraries.

For this reason we divided the comparisons of the present paper in two
different sections. Section 3 studies the flux distributions given by some
model atmosphere grids in order to assess the ability of those models
in predicting broad-band colours. In Section 4 we change our focus to 
libraries that aim at high resolution studies, testing their ability
to reproduce higher resolution spectral features.
The grids evaluated in the present work are briefly described below.

\subsection{Model atmosphere flux distributions}

Amongst several model atmosphere grids available in literature 
\citep[e.g.][]{kurucz93, hauschildt+96, pauldrach+01, heiter+02, TLUSTY03}, 
we selected three grids that cover a large parameter
space in effective temperatures {\Teff} and superficial gravities {\it log g}: \citet[hereafter ATLAS9]{ATLASODFNEW}, 
\citet[hereafter MARCS]{MARCS05} and \citet[hereafter PHOENIX]{PHOENIX05}.

Based on \citet{kurucz93} codes, the ATLAS9 model atmospheres
follow the classical approximations of steady-state, homogeneous,
LTE, plane-parallel layers that extend vertically through
the region where the lines are formed. 
In its more recent version 
\citep{ATLASODFNEW}\footnote{http://wwwuser.oat.ts.astro.it/castelli/grids.html}, 
${\alpha}_{ML}$ is assumed to be 1.25 to fit the energy 
distribution from the centre of the Sun. All models are computed
with the convection option switched on and with the
overshooting option switched off. The convective flux decreases
with increasing {\Teff} and it naturally disappears for {\Teff} $\sim$ 9000K.
The models are available in the range 3500K $\leq$ {\Teff} $\leq$ 50000K.

Plane-parallel LTE models will fail 
wherever sphericity (specially important
for giant stars) and non-LTE effects (for very hot stars)
are evident. 
Two models that take sphericity into account are PHOENIX
and MARCS. 

PHOENIX \citep{hauschildt+96}
is a multi-purpose stellar model atmosphere code for plane-parallel 
and spherical models. The original versions of PHOENIX were developed
for the modelling of novae and supernovae ejecta 
\citep[][and references therein]{hauschildt+99}.
The most recent grid is presented in 
\citet{PHOENIX05}\footnote{ftp://ftp.hs.uni-hamburg.de/pub/outgoing/phoenix/GAIA}.
The equilibrium of Phoenix is solved simultaneously for 40 elements,
with usually two to six ionisation stages per element and 600
relevant molecular species for oxygen-rich ideal gas compositions.
The chemistry has been gradually updated with additional molecular
species since the original code.
The convective mixing
is treated according to the mixing-length theory, 
assuming ${\alpha}_{ML}$ = 2.0.
Both atomic and molecular lines are treated with direct opacity
sampling method. 
PHOENIX models cover the range 2000K $\leq$ {\Teff} $\leq$ 10000K.

MARCS models have 
undergone several improvements since the original code by 
\citet{gustafsson+75}, the most important ones being the replacement of the 
ODF technique by OS technique, the possibility to use a spherically
symmetric geometry for extended objects, and major improvements
of the line and continuous opacities 
\citep{plez+92}.
The common assumptions of spherical or plane-parallel stratification
in homogeneous stationary layers, hydrostatic equilibrium
and LTE are made. Energy conservation is required for
radiative and convective flux, where the energy
transport due to convection is treated through
the local mixing-length theory by
\citet{henyey+65}.
The mixing-length {\it l} is chosen as 1.5H$_{p}$,
which is a reasonable
quantity to simulate the temperature structure beneath
the photosphere 
\citep{nordlund_dravins90}.
The most recent version of the MARCS grids
is presented in \citet{MARCS05}\footnote{http://marcs.astro.uu.se/}.
The models 
cover 4000 $\leq$ {\Teff} $\leq$ 8000K and adopt plane-parallel geometry 
for the dwarfs (log g $\geq $ 3.0)
and spherical geometry for the giants (log g $\leq $ 3.5; 
both geometries are available for log g values of 3.0 and 3.5).

The three sets of models adopt a micro-turbulent velocity of 2 km s$^{-1}$
and are computed for 1 solar mass.

\subsection{High resolution synthetic spectral libraries}

Amongst the higher resolution synthetic libraries, we selected three
of the most recent ones which are publicly available, 
each of them with an outstanding improvement compared to previous
ones. \citet[][hereafter {\it Munari}]{munari+05}\footnote{http://archives.pd.astro.it/2500-10500/}
has an impressive coverage of the
HR diagram. Their models are based on \citet{kurucz93} codes and ATLAS9 grid, 
covering
2500 $-$ 10500 {\angstron} in wavelength range at a maximum resolution of 
R=20000. 
They range from 3500 K to 47500 K in {\Teff}, with {\it log g} varying 
between 0.0 and 5.0 dex, for different values of metallicity,
$\alpha$-enhancement, rotational velocity and micro-turbulent velocity. 

The library by 
\citet[][hereafter {\it Coelho}]{coelho+05}\footnote{http://www.mpa-garching.mpg.de/PUBLICATIONS/{\par}DATA/SYNTHSTELLIB/synthetic\_stellar\_spectra.html}, 
also based on ATLAS9
model atmospheres, had a 
special care for low temperature stars, employing a detailed 
and calibrated line list that has been improved along the years (see the 
original paper for a list of references). 
Their models cover from 3000 {\angstron} 
to 1.8 $\micron$ spanning from 3500 K to 7000 K, with {\it log g} varying
between 0.0 and 5.0 dex, also covering different metallicities
and $\alpha$-enhancement.

\citet[][hereafter {\it Martins}]{martins+05}\footnote{http://www.astro.iag.usp.br/{$\sim$}lucimara/library.htm}
searched the literature
for the best available codes for each range of temperatures
and used them to build the models. They used 
\citet{hubeny88}, \citet{hubeny_lanz95}, \citet{lanz_hubeny03}
model atmospheres
considering non-LTE for hot stars, ATLAS9 models for intermediate
temperature stars and PHOENIX 
line-blanketed models for very cool stars. The library
covers from 3000 to 7000 {\angstron}, with temperatures ranging
from 3000K to 55000K and {\it log g} from -0.5 to 5.5 dex, for 4 different 
metallicities (but no $\alpha$-enhancement).

\section{Evaluating the Flux Distributions: Broad band colours}

A convenient way of comparing the flux distributions given by the 
model grids with observations
is through broad-band colours, which are likely to be the first observables
expected to be predicted by spectral stellar population models.

In order to do this comparison, we selected 
pairs of {\Teff} and {\it log g} that are representative of an isochrone of 
a young and
an old population (10 Myrs and 10 Gyrs). 
The pairs were selected
to uniformly cover {\Teff}, respecting the spacing of each set of models
(ATLAS9 and MARCS have steps of 250K, and PHOENIX has steps of
200K).
The isochrones adopted are the ones by \citet{girardi+02}, for solar
metallicity composition. The transformation to observed colours
were done adopting the UBVRIJHK empirical calibration by 
\citet[][hereafter WL07]{worthey_lee07}~\footnote{Colour-temperature 
table and interpolation program available at http://astro.wsu.edu/models/}.
In that work, the authors used stars with measured photometry and 
known metallicity [Fe/H] to generate colour-colour relations that include the 
abundance dependence.
They further added colour-temperature relations until the whole
parameter range was covered, taking medians in regions where more than
one relation applied. The colour$-${\Teff} relations were obtained from
several sources in literature, mainly from empirical work, but also
from theoretical work. At both ends of the {\Teff} range, the relations
were taken purely from
empirical sources; in the middle range, the theoretical relations by
VandenBerg \& Clem (2003) for V-I were added, and 
behaved well compared to empirical
ones. Any other theoretical relation employed was used with a
lesser weight (G. Worthey, priv. comm. See also Figures 7 and 8 in
WL07). Therefore, we expect the relations by WL07 to be a close match
to observations, and that the theoretical relations, which could bias
our comparisons, do not have an important weight.

The magnitudes predicted by ATLAS9, MARCS and PHOENIX grids were measured using the
IRAF task {\it sbands}, adopting the filter transmission curves of the photometric
systems adopted in WL07.
Zero point corrections were applied to the model magnitudes using the Vega
model by 
\citet{castelli_kurucz94}\footnote{Available at http://wwwuser.oat.ts.astro.it/castelli/vega.html}, 
and adopting Vega
magnitudes : U$_{Johnson}$ = 0.02, B$_{Johnson}$ = 0.03, V$_{Johnson}$ = 0.03,
R$_{Cousin}=0.039$, I$_{Cousin}=0.035$, J$_{Bessell}= 0.02$, 
H$_{Bessell}= 0.02$, K$_{Bessell}= 0.02$.

\begin{figure*}
\resizebox{!}{12cm}{\includegraphics{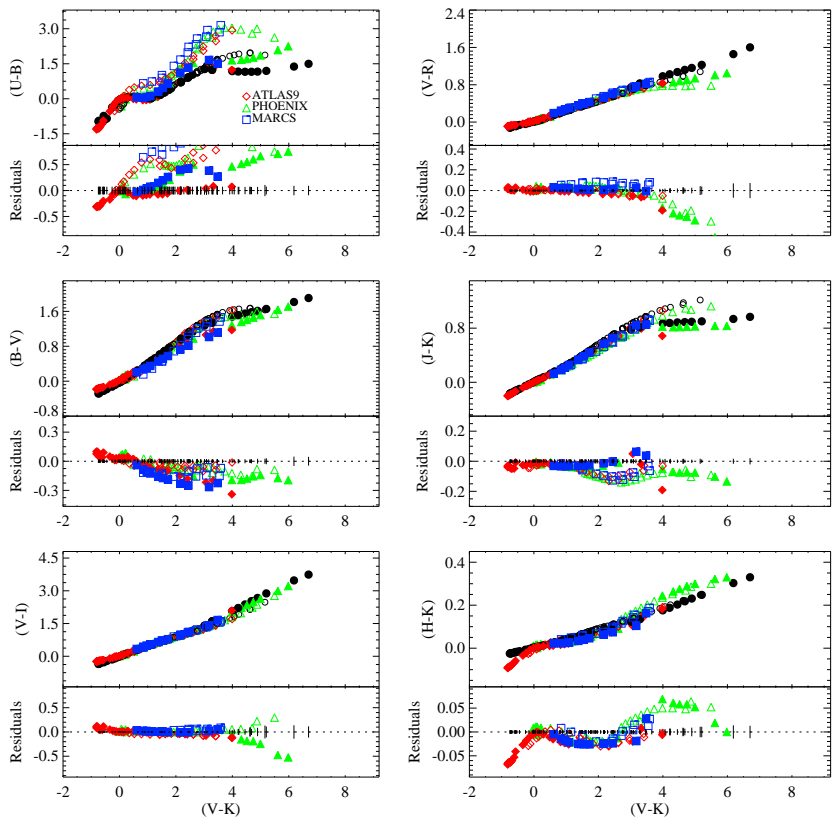}}
\caption{Comparison between the colours predicted by synthetic flux distributions 
and an empirical colour-temperature relation, for stars
representing a 10 Myrs isochrone from \citet{girardi+02}. Red
diamonds correspond to ATLAS9 models, green triangles to
PHOENIX models and the blue squares to MARCS models.
 Filled and open symbols represent dwarfs
(log g $\geq$ 3.0) and giant stars (log g $<$ 3.0), respectively. Circles
are the values expected from the empirical relation of \citet{worthey_lee07}.
On the bottom of each colour plot it is 
shown the residuals (difference between the models and the values
obtained through the empirical calibration). The thin black vertical lines
in this plot represent the error bars of the empirical calibration.
} 
\label{f:cores10My}
\end{figure*}

\begin{figure*}
\resizebox{!}{12cm}{\includegraphics{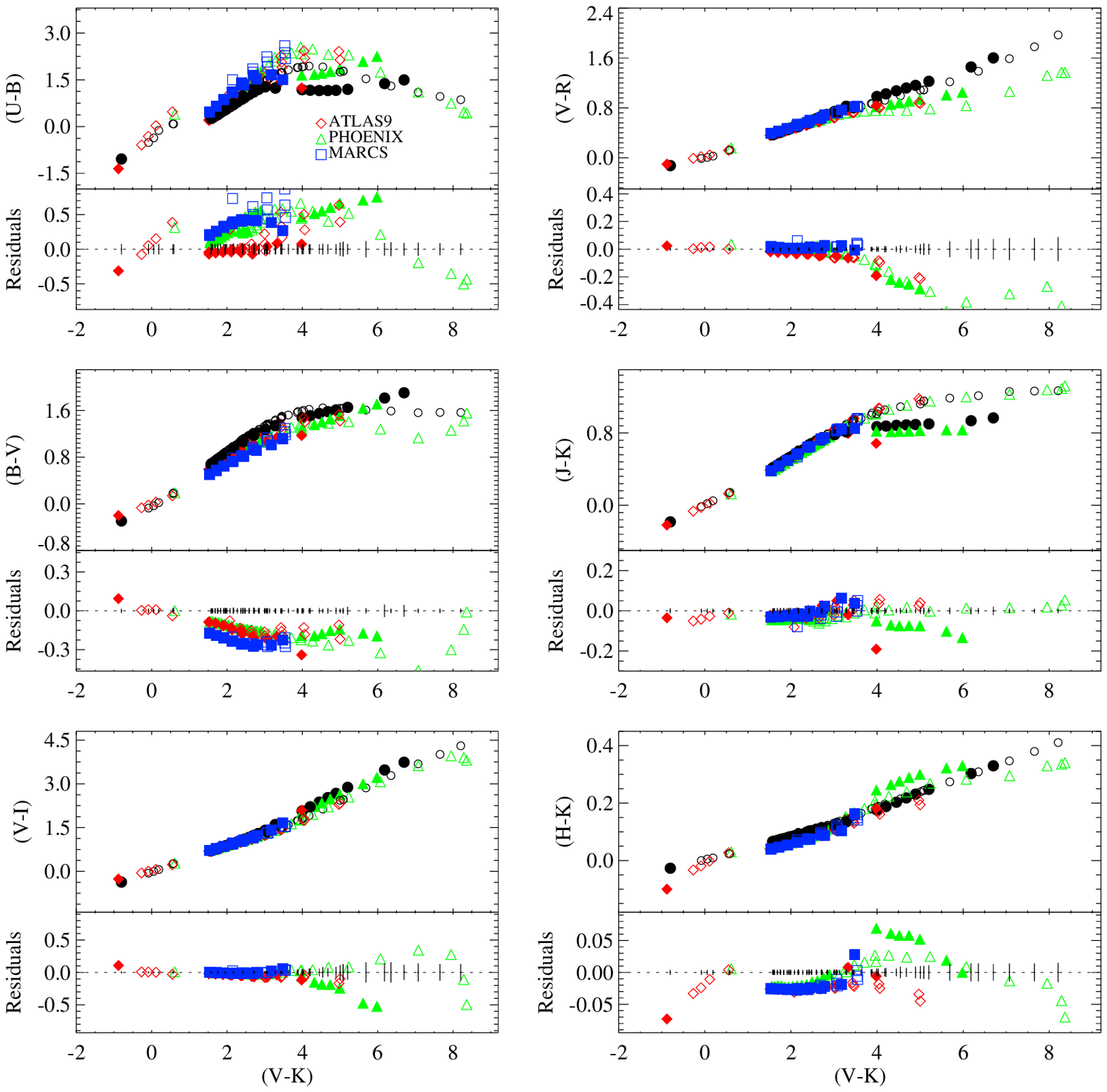}}
\caption{The same as Figure \ref{f:cores10My}, for stars representing a population
of 10 Gyrs.
} 
\label{f:cores10Gy}
\end{figure*}

The comparison between the empirical relation and the model predictions
are given in Figures \ref{f:cores10My} and \ref{f:cores10Gy} for
the 10 Myrs and 10 Gyrs isochrones respectively. The empirical relation
is presented as black circles. ATLAS9 predictions are given in
red diamonds, blue squares are predictions for MARCS models, 
and green triangles for PHOENIX. Filled and open symbols represent dwarfs
(log g $\geq$ 3.0) and giant stars (log g $<$ 3.0), respectively.

The results are presented in colour-colour relations where 
on the $x$ axis is shown the (V-K) colour,
which is a good tracer of {\Teff} (higher values of {\Teff} correspond to lower 
values of V-K). The six panels in each figure show different colours 
in the $y$ axis.
The residuals (model minus empirical) between the 
model colours and the WL07 calibration for each {\Teff}, {\it log g}
pair is shown below each colour-colour panel, where the error bars
indicate the uncertainties of the WL07 calibration.

For stars 4000~K $\leq$ \Teff $\leq$ 8000~K, which is the interval that is common
to all sets of models, we present in Tables \ref{t:res10Myr} and \ref{t:res10Gyr}
the average absolute differences
between model and empirical relations, for the 10 Gyrs and 10 Myrs populations
respectively. 

It can be seen from Figures \ref{f:cores10My} and \ref{f:cores10Gy} 
that the three set of models show a similar behaviour among themselves for
a large range in {\Teff} (V-K), and are a good
reproduction of the empirical relation for the colours V-I, V-R and J-K. 
The residuals are larger for 
cooler stars (V-K $\apprge$ 3), for all colours.
There is a tendency in all models to under-predict the B-V and H-K colours. 
The colour where the models differ more strongly is U-B:
in the case of Figure \ref{f:cores10Gy} (10 Gyrs isochrone), 
we note that in the range 1 $\apprle$ V-K $\apprle$ 3
(which mainly represents the turn-off stars)
ATLAS9 models reproduce considerably better the observations than 
either PHOENIX or MARCS. 
The situation is more complex for the same colour in the young population 
(Figure \ref{f:cores10My}) and all residuals are larger, specially for the giants. 
In the case of the dwarfs, ATLAS9 is still the set of models that best
reproduces the empirical relation.
The differences are typically smaller for the visual colours, 
and for V-I and V-R colours ATLAS9 presents 
on average higher residuals than 
MARCS or PHOENIX, likely due to the different implementations of molecular
opacities.
For the near-infrared colours, the behaviour
is quite similar for the three sets of models.

The reason for the large difference in the U-B colour is unclear to the
present authors. Differences in the 
implementation of both line blanketing and continuum
opacities, and also differences in calibration of the convection treatment
might be playing a role.
The effect of both line blanketing and continuum opacities 
in the near-UV and UV fluxes is a long standing (and sometimes
confusing) problem. Each set
of models has its particular implementation, 
and we refer the reader to 
\citet{houdashelt+00,prieto_lambert00,peterson+01,prieto+03,castelli_kurucz04,garcia-gil+05}
and references therein for appropriate discussions on the subject. 
The effect of the convection treatment on broad-band colours 
have been discussed, for example, in \citet{heiter+02}, and indeed we note
that the three sets of models present different values of the mixing 
length parameter ${\alpha}_{ML}$. However, \citet{kucinskas+05} have shown
that the effect of different ${\alpha}_{ML}$ is not significant, and important
effects appear only when more realistic 3D computations take place.
Nevertheless, they focused their analysis in 
late-type giants, and therefore it remains an open question if different 
${\alpha}_{ML}$ could explain the differences we see here for the 
parameters typical of turn-off stars.

Concerning the cooler stars, usually around V-K $\sim$ 3 ({\Teff} $\sim$ 4250K) the models start to 
deviate from the empirical relation. It is interesting to see that the model predictions are
not strikingly different among the sets of models analysed here (at least for
{\Teff} $\geq$ 3500K), even though
ATLAS9 models are computed in plane-parallel geometry and PHOENIX models in 
spherical geometry (MARCS models present both geometries).
\citet{kucinskas+05} present a very
detailed analysis of broad-band colours for late-type giants, and test
the effect of several model parameters on the broad-band colours predictions 
(namely molecular opacities, micro-turbulent velocities, stellar mass
and treatment of convection). Those authors note that it is possible
that spherical models may not be sufficient, and additional 
effects as convection, variability and mass loss, become increasingly 
important for cooler stars.

\begin{table}
\caption{Mean absolute residuals for the broad-band colours. These
values were obtained for the 10 Myrs isochrone and for
the interval 4000K $\leq$ \Teff $\leq$ 8000K.}
\label{t:res10Myr}
\begin{center}
\begin {tabular}{lcccc}
\hline
Colour & ATLAS9  & MARCS 	& PHOENIX & Mean error \\
\hline
U$-$B& 0.370 &  0.695 & 0.611 & 0.073 \\
B$-$V& 0.070 &  0.145 & 0.066 & 0.020 \\
V$-$I& 0.041 &  0.029 & 0.010 & 0.015 \\
V$-$R& 0.022 &  0.045 & 0.026 & 0.012 \\
J$-$K& 0.049 &  0.056 & 0.079 & 0.013 \\
H$-$K& 0.018 &  0.016 & 0.019 & 0.004 \\
\hline
\end{tabular}
\end{center}
\end{table}

\begin{table}
\caption{Mean absolute residuals for the broad-band colours. These
values were obtained for the 10 Gyrs isochrone and for
the interval 4000K $\leq$ \Teff $\leq$ 8000K.}
\label{t:res10Gyr}
\begin{center}
\begin {tabular}{lcccc}
\hline
Colour & ATLAS & MARCS & PHOENIX & Mean error \\
\hline
U$-$B& 0.105 & 0.440 & 0.309  & 0.073\\
B$-$V& 0.146 & 0.235 & 0.126  & 0.020\\
V$-$I& 0.048 & 0.015 & 0.009  & 0.015\\
V$-$R& 0.038 & 0.017 & 0.016  & 0.012\\
J$-$K& 0.023 & 0.027 & 0.034  & 0.013\\
H$-$K& 0.024 & 0.022 & 0.018  & 0.004\\
\hline
\end{tabular}
\end{center}
\end{table}

\section{Evaluating the high resolution features: Spectral indices}

A convenient way to evaluate the theoretical spectra 
is to measure widely used spectral indices and compare them 
with the observed values.
This approach will not evaluate the quality of the model spectrum at its 
full wavelength coverage, but allows a presentation of the results in a scale that is
familiar to the user of stellar population models.

We compared {\it Coelho}, {\it Martins} and {\it Munari} libraries with three of the 
most complete empirical libraries available: Indo-US, MILES and
Elodie. 

\subsection{Overview of the Empirical Libraries}
The first empirical stellar library that provided flux
calibrated spectra was presented in \citet{jones98}. With
moderately high resolution (1.8 {\angstron}), this library was used by
\citet{vazdekis99} to produce for the first time spectral stellar population 
models at high resolution. However, {\it Jones} library is limited
to two narrow wavelength regions (3820-4500 {\angstron} and 4780-5460 
{\angstron}), and
it's sparse in dwarfs hotter than about 7000 K and metal-poor giants.

STELIB\footnote{http://www.ast.obs-mip.fr/users/leborgne/stelib/index.html} \citep[]{stelib}
represents a substantial improvement 
over previous libraries. It
consists of 249 stellar spectra in the range of 3200 {\angstron}
to 9500 {\angstron}, with an spectral resolution of about 3{\angstron} 
(R=2000). This is the base library for the
widely used \citet{BC03} stellar population models.

Following this work, \citet{INDOUS} published
Indo-US \footnote{http://www.noao.edu/cflib}, 
a library with resolution down to FWHM $\sim$ 1{\angstron} and a 
good coverage of the colour magnitude diagram. Indo-US has a much higher
number of stars (1273),
with spectra ranging from 3460 {\angstron} to 9464 {\angstron}.
They cover a fair range in atmospheric parameters. The main concern on this
library regards its spectrophotometry, which was obtained by 
fitting each observation to a standard spectral energy distribution with 
a close match in spectral type, using the compilation of 
\citet{pickles98}.

\citet{ELODIE} published the ELODIE Library\footnote{http:
//www.obs.u-bordeaux1.fr/m2a/soubiran/elodie\_library.html},
which has been updated since then. In its 
current version (Elodie.3) there are 1388 starts, in the wavelength range 4000 
to 6800
{\angstron}. Although it has a more limited wavelength coverage with 
respect to the others, it has a very high spectral resolution
(R=10000 for flux calibrated spectra and R=42000 for flux normalised
to the pseudo-continuum). But the flux calibration
of this library might be compromised by the 
use of an echelle spectrograph. 

Another library that became available recently is
MILES\footnote{http://www.ucm.es/info/Astrof/miles/miles.html}
\citep{MILES,cenarro+07}. The spectra ranges
from 3525 {\angstron} to 7500 {\angstron}, at a 2.3 {\angstron} (FWHM) 
resolution. 
This library, with 985 stars, was carefully
created trying to fill the major gaps that existed in other
empirical libraries.

The Next Generation Stellar Library (NGST, Gregg et al. 2004) is
yet another library soon to be publicly available,
which is an UV/optical (from 1660 to 10200 {\angstron}) stellar spectral atlas 
using STIS-HST (PID 9786).
The advantage of this library is that, being obtained with
STIS at Hubble Space Telescope, it presents an unprecedented internally consistent
flux calibration across all wavelengths.

\begin{figure}
\resizebox{\columnwidth}{!}{\includegraphics{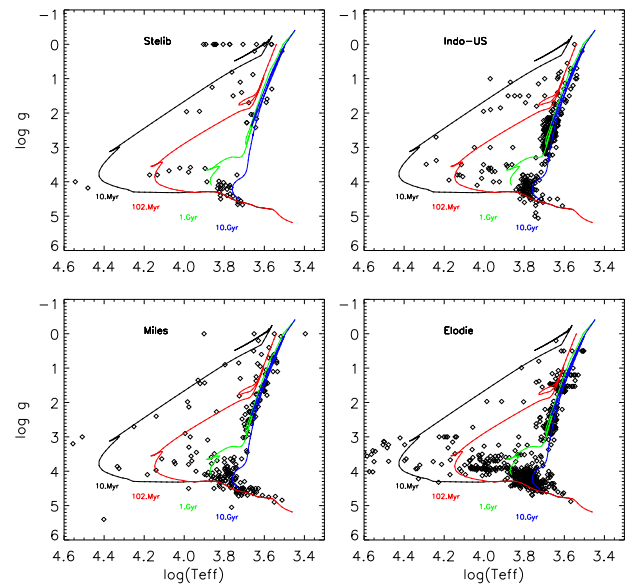}}
\caption{Distribution of stars with solar metallicity in four
empirical libraries. The solid lines are the 
solar metallicity isochrones by \citet{girardi+00}
for four different ages: black is 10Myrs, red is 100Myrs,
blue is 1Gyrs and red is 10 Gyrs.
} 
\label{f:emp_lib}
\end{figure}

Figure \ref{f:emp_lib} shows the coverage in temperature and gravity
of four empirical libraries
(STELIB, Indo-US, MILES and ELODIE), overplotted on
isochrones from
\citet{girardi+02}
for ages 10~Myrs, 100~Myrs, 1~Gyrs and 10~Gyrs.
All isochrones are for solar metallicity, which is the regime where
the empirical libraries are more complete. The stars plotted are the ones
with metallicity in the range -0.1 $\leq$ [Fe/H] $\leq$ 0.1.
It is clear that libraries have been improving in terms of parameter coverage,
but this is a hard task and some holes still exist. Hot stars are 
missing
in most of the libraries, being really sparse towards O and B
stars.
Libraries tend to have the majority of stars for temperatures 
between 5000 K
and 7000 K and there is a sudden drop for lower temperatures, specially below 
4000 K. 
MILES has the best coverage for 
lower temperatures, while ELODIE is the most complete in the high
temperature end. STELIB has only one O star, and only one dwarf below 4000K.
Indo-US has no stars with {\Teff} $>$ 27000K, and no dwarf below 4000K.

\subsection{Spectral indices comparison}
A total of thirty-five spectral indices were measured in the spectra
of the three high resolution synthetic libraries to be studied 
({\it Coelho}, {\it Martins} and {\it Munari}), and on three of the empirical
libraries (Indo-US, ELODIE and MILES).
We selected all the Lick/IDS indices as defined by
\citet{worthey+94} and \citet{worthey_ottaviani97}.
We extended the wavelength coverage adding the index D4000 \citep{balogh+99},
some indices defined in \citet{serven+05} and the near infrared indices by 
\citet{diaz+89}. The indices cover the wavelength region 
$\sim$ 3500{\angstron} to $\sim$ 8700{\angstron} and 
are summarised in the Appendix (Table A1). All the indices except D4000 are defined
by a central bandpass
bracketed by two pseudo-continua bandpasses, which are used
to draw a straight line to determine the continuum level at
the feature. Atomic features are expressed in angstroms and
molecular features in magnitudes.
D4000 is defined using 100 {\angstron}
continuum bandpasses to measure the break (3850 - 3950 {\angstron}
and 4000 - 4100 {\angstron}). 

We compared each synthetic library to each empirical library.
For each empirical library, we selected all stars with metallicity in the range
-0.1 $\leq$ [Fe/H] $\leq$ 0.1. For each
star, the closest model in {\Teff} and {\it log g} ([Fe/H] = 0.0) was 
selected in each of the synthetic libraries.
The closest model was chosen based on the smaller distance (d)
to the {\Teff} $\times$ {\it log g} plane, defined in equation
\ref{eq:dist}, where {\Teff} and log g are parameters of the models, and T$_{obs}$ and 
(log g)$_{obs}$ are parameters of the empirical libraries.

\begin{center}
\begin{equation}
\label{eq:dist}
d=\sqrt{{\left(\frac{T_{\rm eff} - T_{obs}}{T_{obs}}\right)}^2+{\left(\frac{log\ g - (log\ g)_{obs}}{(log\ g)_{obs}}\right)}^2}
\end{equation}
\end{center}

The typical parameter spacing of the models (250K in {\Teff} 
and 0.5 dex in {\it log g}) is of the same order 
of the accuracy of the atmospheric
parameters in the empirical libraries. Therefore, we believe the 
closest model is a reasonable approach. 
The theoretical libraries were degraded to the resolution of each empirical
library prior to the measurements of the indices. The exception was the ELODIE
library, whose superior resolution could only be matched by {\it Coelho} library. 
In this case the theoretical libraries and ELODIE were degraded to a common
resolution of FWHM = 0.3{\angstron}.

Figures for all the comparisons
are presented in the Appendix (on-line material). 
Figures \ref{f:idx_balmer} to \ref{f:idx_tio} show the results for
some of the  indices. 
The data points on the figures are the median values for 
each {\Teff} and {\it log g}
bin in the empirical libraries, 
and the error bars are the correspondent one sigma dispersion of the 
{\it empirical} measurements for that parameter bin. 
A point with no error bar implies that there was only one star for that 
{\Teff}
and {\it log g} bin. We colour coded the stars in three {\Teff}
intervals:
blue squares are stars with {\Teff} $>$ 7000K, green diamonds
are stars with 4500K $<$ {\Teff} $\leq$ 7000K, and red asterisks are stars with
{\Teff} $\leq$ 4500K. The black crosses are stars with {\Teff} $<$ 3500 K, but they 
are
really rare.
We also separated them by gravity: dwarf stars (log g $\geq$ 3.0) are 
represented by filled symbols and giant stars (log g $<$ 3.0) are
represented by open symbols.
The black line in each plot shows 
the one to one relation.
The thick black symbols indicate the location 
of a Sun-like dwarf (cross; {\Teff} = 5750K and {\it log g} = 4.5), and a
typical K1 giant (diamond; {\Teff} = 4250K and {\it log g} = 1.5). 
The K1 giant have all parameters but metallicity close to the star Arcturus.
We show the position of these particular stars on the plots because line lists are 
usually calibrated based on their high resolution spectra. 
Also shown
in each plot is the {\it adev} value for each temperature range,
a statistical measurement of how much
each model is representing the stars in that range.
$Adev$ takes into account the distance of each theoretical point from
the one-to-one line in the index plots, and is defined as:

\begin{center}
\begin{equation}
adev=\frac{1}{N} \sum \left|\frac{(I_t - I_e)}{I_e}\right|
\end{equation}
\end{center}

where N is the number of stars,
I$_t$ is the measure of the index on the theoretical library
and I$_e$ is the measure of the index on the empirical library.

First thing to notice in these plots is that the error bars are non-negligible, specially
for the low temperature stars. This is a consequence of the large
uncertainties in the atmospheric parameters of
these stars. The determination of those parameters in cool stars is known to be 
a real challenge.
For the high temperature stars it is clear that
the spread between each point is very small for most of the indices. This
is somewhat expected, since there are fewer metallic lines as you go 
up in temperature, and therefore many of these indices will give  
essentially no information in this case. 
 
We organised the analysis grouping the results in four categories, 
related to the chemical species that dominate the index.
It is worth remember that
no index is sensible to only one element \citep[see e.g. tables at][]{serven+05},
but we attempted to categorise the indices by its main element.

{\it Balmer lines:}
Include the indices H$\beta$, H$\gamma_A$ and H$\delta_A$. 
In general the hydrogen
indices are well reproduced by all models down to 4500K. For the very
low temperature stars, models start to deviate from observational
libraries, clearly subestimating the indices, as shown in 
Figure \ref{f:idx_balmer} for H$\gamma_A$. 
It is known that Hydrogen lines computed in LTE match well the 
wings, 
but cannot reproduce the core of the lines. Fine tuned micro turbulence 
velocities or mixing length to pressure scale height ratio 
$\ell$/H$_{\rm p}$ 
were suggested in literature to improve the match in the solar spectrum
\citep[e.g.][]{fuhrmann+93,vantveer+96}, but the same
parameters would not necessarily improve the  results for other
spectral types. 
A realistic match would 
require NLTE computations of H lines, only available for very hot stars. 
Besides, the bottom of the hydrogen lines form in the chromosphere, not included
in the model atmospheres grids. 
Another point to note is that although theses indices are aimed at measuring
H lines, in low temperature stars the actual hydrogen
lines are considerably weak, and the metallic lines
can be dominating the index.
In this case, it is not clear if the main reason why
the models are failing in reproducing the observed values is because of
the non-satisfactory line core modelling, or because the dominance of 
uncalibrated metallic lines.

\begin{figure*}
\resizebox{!}{10cm}{\includegraphics{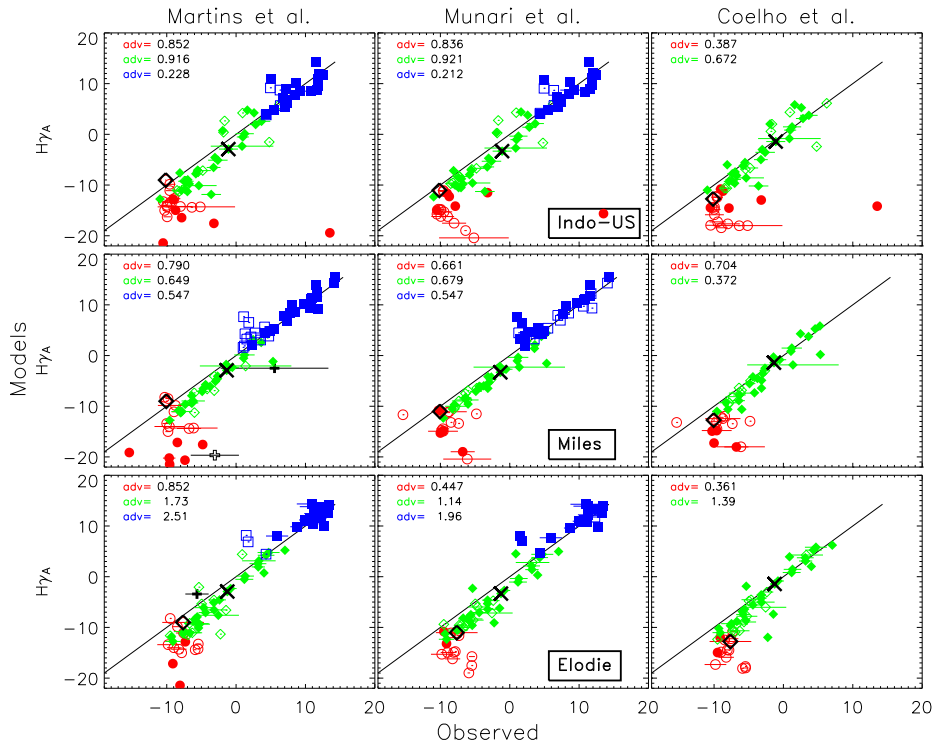}}
\caption{Comparison of the index H$\gamma_A$ measured in the empirical and 
theoretical libraries. Different symbols and colours represent three intervals of 
temperature: blue squares are stars with Teff $<$ 7000K, green diamonds
are stars with 4500K $<$ Teff $\leq$ 7000K and red circles are stars with
Teff $\leq$ 4500K. Filled and open symbols represent dwarfs
(log g $\geq$ 3.0) and giant stars (log g $<$ 3.0), respectively.
The black crosses are stars with Teff $<$ 3500 K. The solid
line is the one to one relation. The thick black symbols 
represent a Sun-like dwarf (cross) and an Arcturus-like giant (diamond).
} 
\label{f:idx_balmer}
\end{figure*}

{\it C and N indices:}
Include the indices CNO3862, CN1, CN2 and G4300. According to \citet{TB95} 
calculations, the indices Ca4227 and Fe4668 are also highly sensitive to 
Carbon abundance
variations, and therefore these two indices could be possibly included in this group.
>From these indices, the sub-sample that is sensitive to both C and N abundances
(CNO3862, CN1, CN2) show significant larger error bars, but the overall 
behaviour seem to be well matched by the 
models. 
Figure \ref{f:idx_CN}, 
that shows the CN$_2$ index, illustrates this effect.
On the other hand, indices that are mainly sensitive to C abundance 
variations (G4300, Ca4227
and Fe4668) systematically deviate from the one to one line for stars cooler
than {\Teff} = 4500K.
Figure \ref{f:idx_G4300} shows the G4300 index, which measures the
G-band of CH at 4300{\angstron}. 
One possible reason for this effect is that 
the C and N abundances relative to Fe were assumed to be solar for all 
synthetic stars, 
but it is well known that the CNO-cycle lowers the C abundance and enhances
the N abundance in giants \citep[e.g.][]{iben67,charbonnel94}. The same effect
on the indices CN$_{1}$ and CN$_{2}$ would not be so clearly seen if the 
variations of C and N somewhat compensate each other.
Nevertheless, 
we could not clearly attribute all the differences in these indices to the
un-modelled CNO mixing.
If the modelling of the CNO cycle was to be
the only or major problem affecting the cool giants, we would
expect the dwarfs (filled symbols; see e.g. Figure \ref{f:idx_G4300}) 
to be a closer match to the observations 
than the giants (open symbols). 
This is not the case, both presenting similar
patterns.
Interestingly, for temperatures
between 4500K and 7000K, {\it Coelho} models reproduces considerably better 
the observations, while the cool end deviates more strongly
than the other synthetic libraries. 
This is probably because the CH lines adopted by 
{\it Coelho} models were computed
with LIFBASE code \citep{lifbase} while {\it Martins} and {\it Munari} models adopt
\citet{kurucz93} molecular lines. This is
a first indicative of how working on the line lists might impact the
model results in significant ways. 

\begin{figure*}
\resizebox{!}{10cm}{\includegraphics{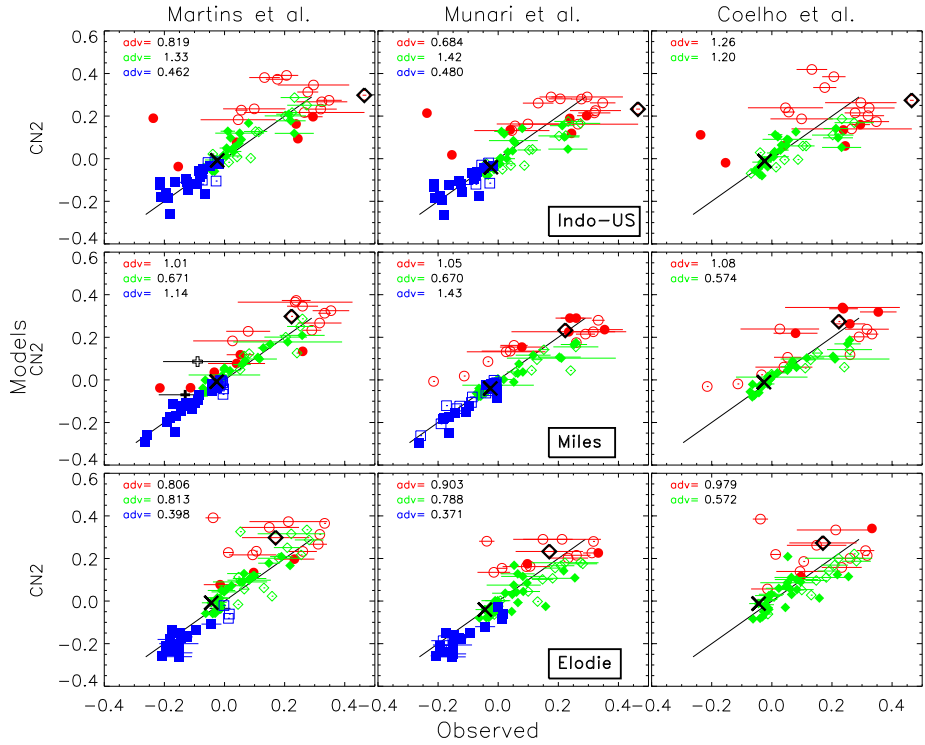}}
\caption{Comparison of the index CN$_2$ measured in the empirical and 
theoretical libraries. This index measures the strength
of the CN$\lambda$4150 absorption band, in magnitudes. 
Symbols and colours are the same as in Figure \ref{f:idx_balmer}.
} 
\label{f:idx_CN}
\end{figure*}

\begin{figure*}
\resizebox{!}{10cm}{\includegraphics{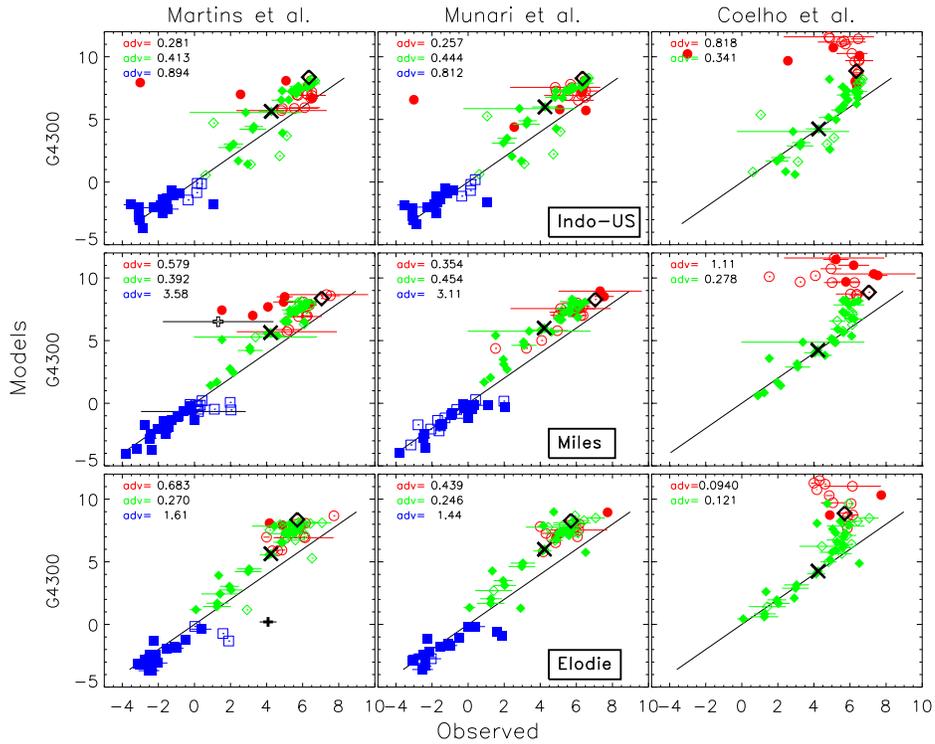}}
\caption{Comparison of the index G4300 measured in the empirical and 
theoretical libraries. Symbols and colours are the same as in Figure \ref{f:idx_balmer}.
} 
\label{f:idx_G4300}
\end{figure*}

{\it Iron peak elements: }
Many of the iron indices are good examples suggesting that 
working on the line lists might
improve the model results significantly. 
Figure \ref{f:idx_fe} shows the behaviour of the index Fe4383, 
where this effect is evident. 
{\it Martins} and {\it Munari} models have similar
line lists, modified from the same \citet{kurucz93} original work, 
while {\it Coelho} models employed its 
independent line list, based on high resolution stellar spectroscopy studies.
The effect of the different line lists is clearly seen.

\begin{figure*}
\resizebox{!}{10cm}{\includegraphics{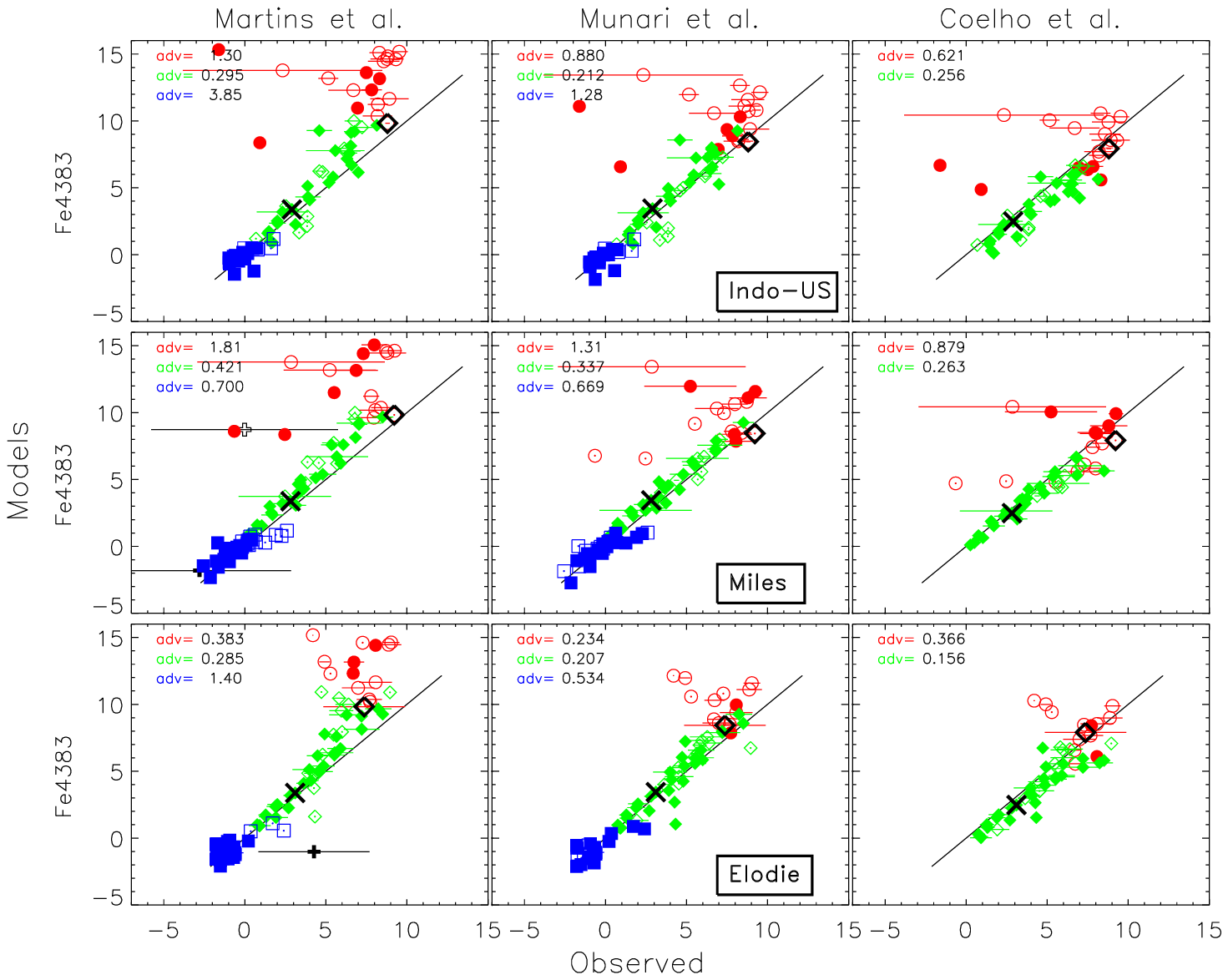}}
\caption{Comparison of the index Fe$\lambda$4383 measured in the empirical and 
theoretical libraries. Symbols and colours are the same as in Figure \ref{f:idx_balmer}.
} 
\label{f:idx_fe}
\end{figure*}

{\it $\alpha$ elements:}
Include all the indices sensitive to Mg, Ca, Ti and O.
In this case there is not a general pattern.
Figure \ref{f:idx_mg} shows the Mg$_2$ index where the
line list from {\it Coelho} reproduces 
significantly better the observed values, specially 
in the low temperature regime. But it is interesting to point out
that for stars cooler than {\Teff} $\sim$ 4250K, this index is 
heavily contaminated
by TiO $\alpha$ molecular features 
\citep[see Figure 13 in][]{coelho+05}.
The Calcium and TiO indices, on the other side, are examples of how
things can be complex. 
Figure \ref{f:idx_ca} shows the index Ca4455.
{\it Coelho} models tend to predict slightly lower values than the observed. 
{\it Munari} models seem to show the same trend, to a lower level. At first
order we could conclude that both models are under-predicting this index, but
\citet{bensby+05} 
studied F and G dwarfs from the thin and thick disc of our galaxy
and found that the [Ca/Fe] tend to be slightly super-solar for stars with 
[Fe/H] solar. In the likely case that the stars in the 
empirical libraries show a 
similar behaviour than the one found by \citet{bensby+05}, 
we should not 
expect the models, calculated with solar
mixture ([Ca/Fe]=0), to actually match the observations. In this case, 
the behaviour of
both {\it Coelho} and {\it Munari} models are consistent with the observations. 
{\it Martins} models 
show a more
complex behaviour: intermediate temperature stars, which were computed 
with SPECTRUM synthesis 
code and line lists and ATLAS9 models, are overestimated; low 
temperature stars, calculated with PHOENIX models and line lists, are 
underestimated.
Figure \ref{f:idx_tio} shows the TiO$_2$ index. This index has no meaning for 
stars with temperatures higher than $\sim$ 4500K, where there is no
TiO in the spectrum to be measured. 
For lower temperature stars the values raises rapidly, 
being extremely sensitive to temperature. This implies that
uncertainties in the {\Teff} adopted for stars in the empirical libraries 
(usually considerably higher 
for low temperature stars) make the comparison with models hardly reliable. 
Given the large uncertainties, models are not failing completely to reproduce
this index.

\begin{figure*}
\resizebox{!}{10cm}{\includegraphics{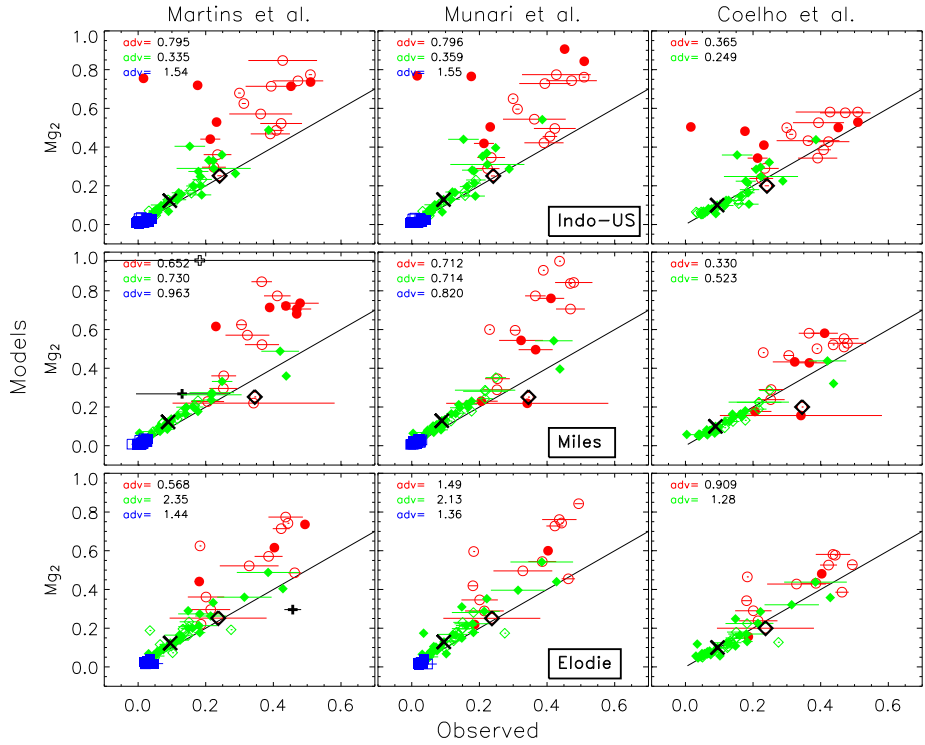}}
\caption{Comparison of the index Mg$_2$ measured in the empirical and 
theoretical libraries. Symbols and colours are the same as in Figure \ref{f:idx_balmer}.
} 
\label{f:idx_mg}
\end{figure*}

\begin{figure*}
\resizebox{!}{10cm}{\includegraphics{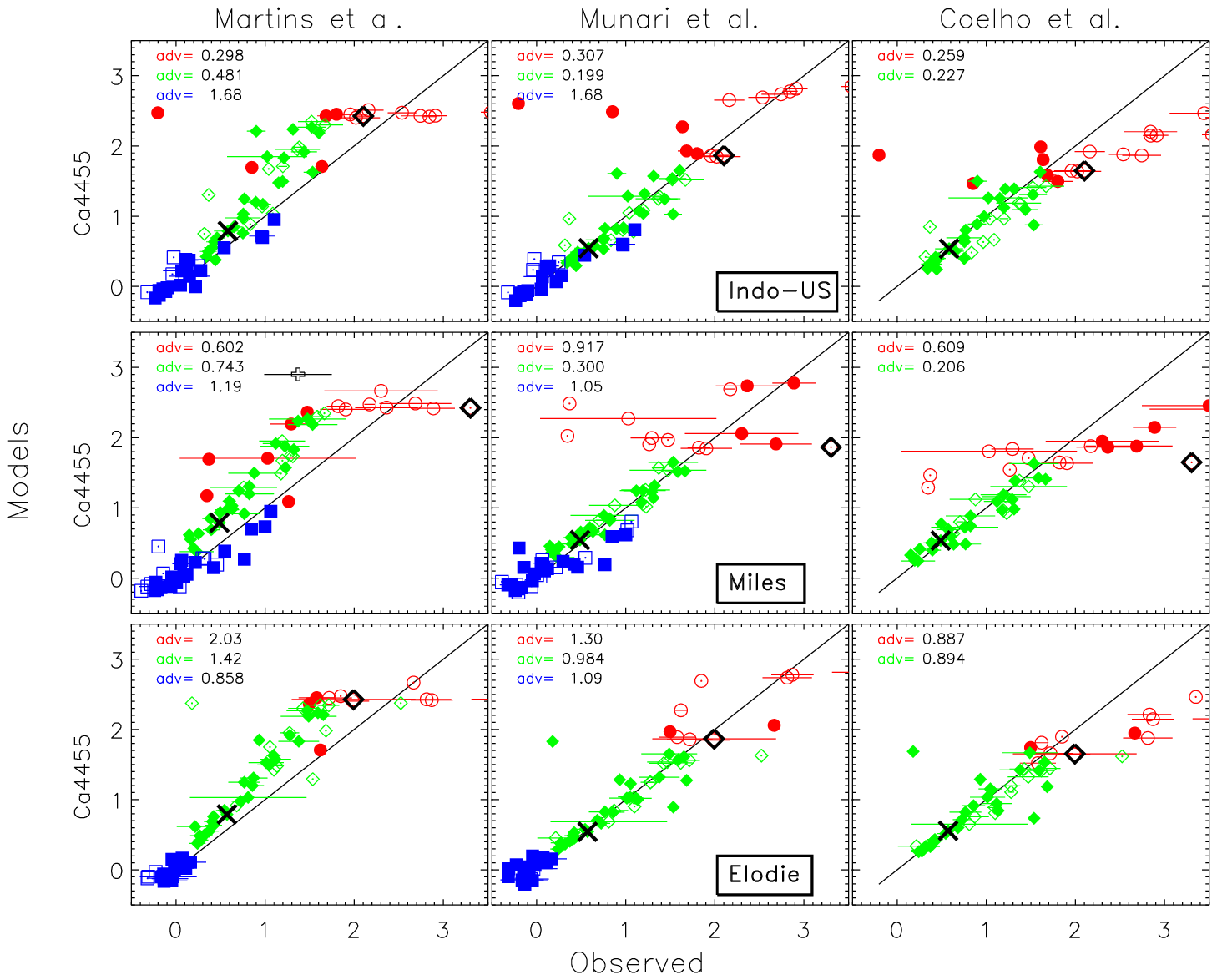}}
\caption{Comparison of the index Ca4455 measured in the empirical and 
theoretical libraries. Symbols and colours are the same as in Figure \ref{f:idx_balmer}
} 
\label{f:idx_ca}
\end{figure*}

\begin{figure*}
\resizebox{!}{10cm}{\includegraphics{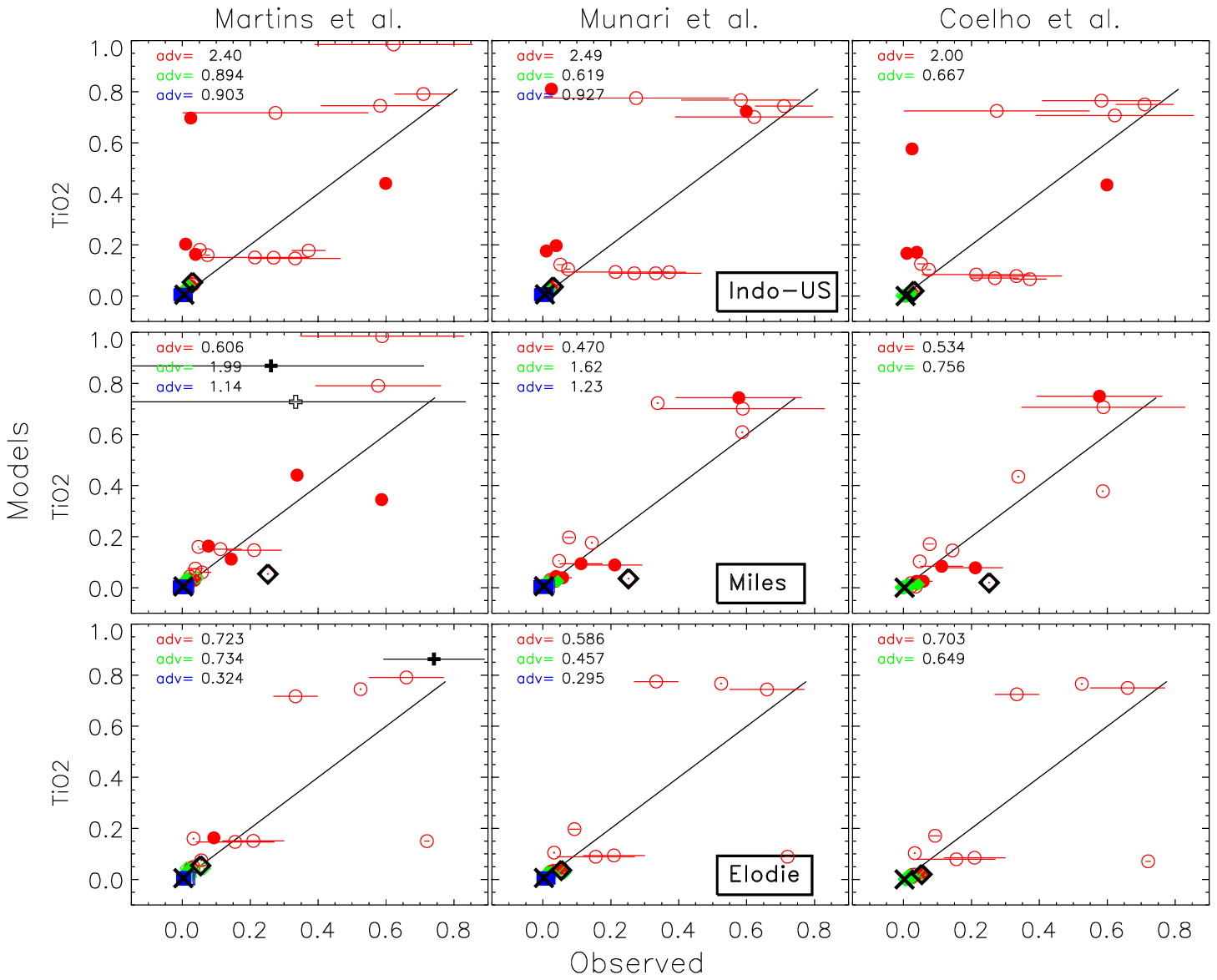}}
\caption{Comparison of the index TiO$_2$ measured in the empirical and 
theoretical libraries. Symbols and colours are the same as in Figure \ref{f:idx_balmer}.
} 
\label{f:idx_tio}
\end{figure*}

\subsection{Dependence on the atmospheric parameters}
It is worth to keep in mind that errors on the empirical libraries,
the most important one being uncertainties in the atmospheric parameters, 
hamper the comparison with the models. 

\begin{figure}
\resizebox{\columnwidth}{!}{\includegraphics{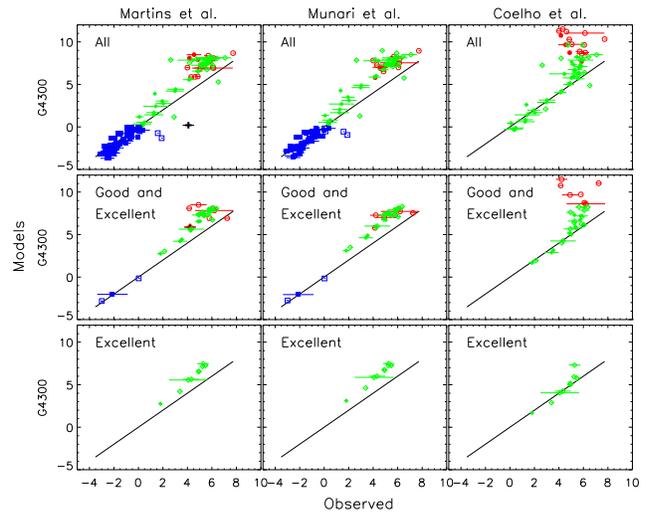}}
\caption{Comparison of the index G4300 measured on the ELODIE library,
filtering by the accuracy flags. First line has all the stars, second line
shows only stars with {\it good} and {\it excellent} atmospheric parameters,
and the third line only stars with {\it excellent} flags. 
Symbols and colours are the same as in Figure \ref{f:idx_balmer}.} 
\label{f:elodie_g4300}
\end{figure}

\begin{figure}
\resizebox{\columnwidth}{!}{\includegraphics{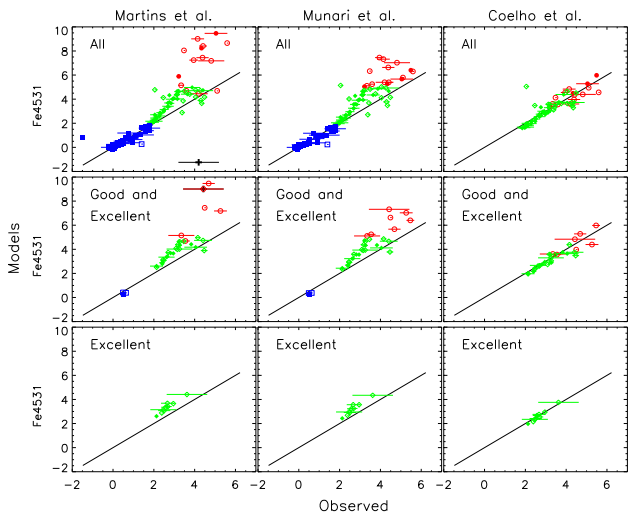}}
\caption{The same as Figure \ref{f:elodie_g4300} for the index Fe4531.
} 
\label{f:elodie_fe4531}
\end{figure}

ELODIE library is the only of the empirical libraries that provides, for each
star, a flag that indicates the accuracy of each atmospheric parameter. 
In order to evaluate how much the accuracy 
might affect our comparisons, Figures \ref{f:elodie_g4300} and 
\ref{f:elodie_fe4531} show the same comparisons as before for the indices 
G4300 and Fe4531, but filtering the observed stars by the quality flag 
of the atmospheric parameters.
On the first line of the figures all stars are plotted. 
On the second line, only stars with {\it good} and {\it excellent} 
flags for the atmospheric parameters. On the third line, 
only the ones with {\it excellent} 
determination. It is clearly noticeable how much the agreement between models and 
observations
can change, based only on stars with very good parameter determinations. 
The drawback, on the other hand, is that this filter limits drastically the 
number of points. 

\subsection{Dependence on the flux calibration}

A second issue that can complicate the comparison between model and
observations are related to flux calibrations uncertainties.
One of the advantages of using spectral indices is that they were designed to be,
as much as possible, insensitive to flux calibration issues. 
That implies that when using these indices to study the properties 
of stellar populations, the continuum shape is not
used to extract information from the spectra. This is particularly useful
when it is not possible to accurately flux calibrate
the observations. 

In order to test how sensitive to flux calibration issues are the indices studied
here, we employed a
modified version of {\it Coelho} library. As explained 
in \S 2, a library focused on spectroscopic use is not suitable to predict broad-band 
colours because it does not generally include the full line blanketing. 
As the libraries stand now, our note to the stellar population modeller $-$ which might be interested
in using any of the synthetic libraries currently available $-$
is that one has to find a compromise between a
library which is good for spectrophotometric predictions or one which is good
for spectroscopic studies. Until the accuracy of the predicted energy levels lines 
is significantly improved \citep[see e.g.][]{kurucz06}, 
the only way of achieving reasonable predictions for both broad-band colours and high
resolution indices is by correcting the pseudo-continuum of current high resolution libraries 
to better match observed colours.
In order to use the high resolution library to build stellar population
models, \citet[]{coelho+07} applies a correction to the original 
library presented in \citet{coelho+05} in order
to compensate for the mentioned missing line opacity. 
In a few words, this correction was done by comparing each star in {\it Coelho} library
to the correspondent flux distributions by ATLAS9 grid.
Smooth corrections to the continuum shape were applied to the stars in {\it Coelho} library
in order to better match the continuum shape of its correspondent flux distribution by ATLAS9.
Therefore, the modified {\it Coelho} library kept the high resolution features of the original 
library, but presents a flux distribution which is closer to that 
predicted when including all blanketing (ATLAS9).

\begin{figure}
\resizebox{\columnwidth}{!}{\includegraphics{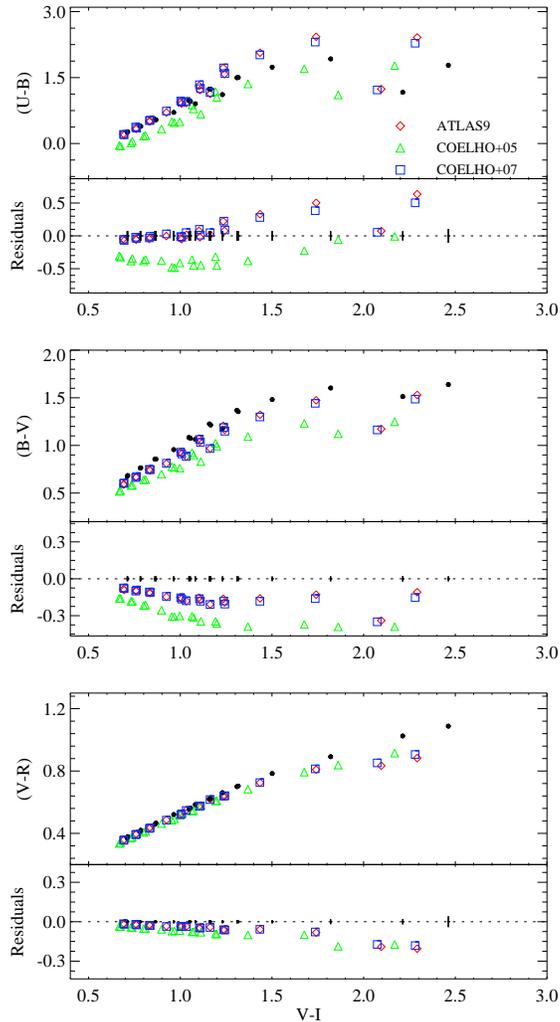}}
\caption{Comparison between the colours predictions from two versions of {\it Coelho} library, 
with and without the empirical correction of the continuum as described in \S 4.4 (blue squares
and green triangles respectively). Red diamonds are the predictions by ATLAS9 models, for comparison.
} 
\label{f:corespau}
\end{figure}

The effect of this correction is shown in 
Figure \ref{f:corespau}, in a similar fashion of the broad-band colours figures at
\S 3. ATLAS9 flux distributions are
shown as red diamonds, {\it Coelho} original library stars are shown as green triangles, 
and the blue squares are the flux corrected stars (the modified {\it Coelho} library 
by \citealt{coelho+07}).The effect of the missing line opacity is clear, specially 
in the blue part of the spectrum.

The spectral indices were then measured in the modified {\it Coelho} library
and compared to the original measurements. These comparisons
can show how smooth changes in the stellar pseudo-continuum can affect the 
measurement of the indices used in the present work. 
As expected, for most of the indices the differences 
between the two measurements
are smaller than 3$\%$. Among the 
classical Lick indices, only Ca4455 and Mg$_1$ are slightly
more sensitive ($\sim$ 5$\%$).

\begin{figure}
\resizebox{\columnwidth}{!}{\includegraphics{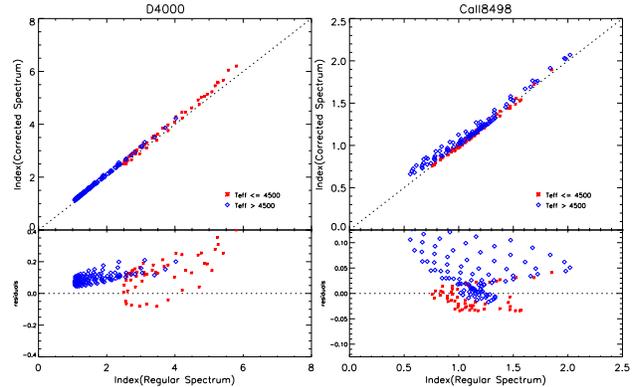}}
\caption{Comparison between indices calculated for two versions of {\it Coelho} library,
with and without the flux correction due to missing line opacity. 
} 
\label{f:fluxtest}
\end{figure}

The notable exceptions are the indices D4000 and the three Ca indices 
in the near infrared, that showed a considerable sensitivity to the 
modifications of the continuum shape (reaching above 10$\%$ in the most
extreme cases).
In Figure \ref{f:fluxtest} we show
the comparisons between the indices calculated with the original library ($x$ axis)
and the flux corrected one ($y$ axis), and the residuals in the bottom panels.
This high sensitivity of D4000 index to flux
calibrations issues has also been noticed by G. Bruzual, V. Wild \& S.
Charlot (priv. comm.)

\subsection{The profile of the H lines in high temperature stars}

Balmer lines play a crucial role in the quantitative spectral analysis
of hot stars. The Stark broadened wings depend on the photospheric 
electron density and, consequently, the stellar gravity {\it log g}.
The line cores on the other hand are more sensitive to the effective
temperature {\Teff}. Thus, the complete Balmer line profiles contain
information about both fundamental atmospheric parameters, {\Teff} and
{\it log g}. The effects of NLTE were demonstrated to be of drastic 
importance since the pioneering work of \citet{auer_mihalas72}, and
have to be considered in order to reproduce these lines. 
\citet{martins+05} already showed that this effect becomes more
important with increasing {\Teff}, making a real
difference for O and early B stars.

\begin{figure}
\resizebox{\columnwidth}{!}{\includegraphics{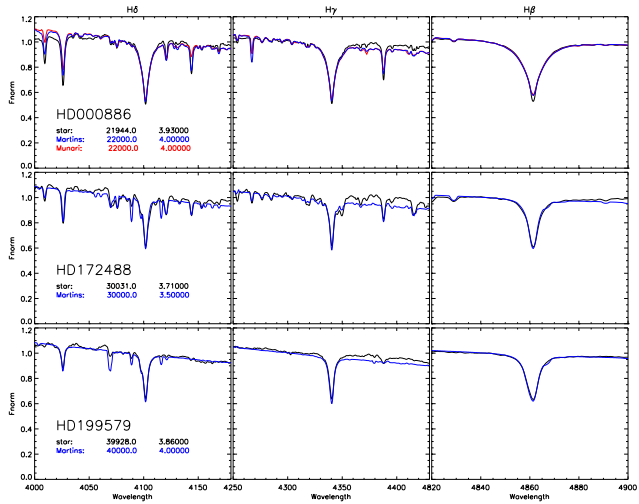}}
\caption{Comparison between models and observations for three of the balmer line profiles. 
Observations
are shown in black, and models are given in blue ({\it Martins}) and red ({\it Munari}).
The star identification and stellar parameters are shown in the plots.
} 
\label{f:blines}
\end{figure}

Figure \ref{f:blines} shows a comparison between three hot stars from
the ELODIE library (which is more complete for hot stars) and 
the theoretical libraries from {\it Martins} and {\it Munari} ({\it Coelho} library
stops at 7000K). The hot stars in {\it Munari} library are also limited to {\it log g} equal to 
4.5 or 5.0, 
while in the empirical libraries the hotter stars have 
3.5 $\leq$ {\it log g} $\leq$ 4.0.
The top line of the figure shows three Balmer lines for a star with {\Teff} $\sim $21000K. 
In this case, both models are LTE. On the H$\beta$ profile this might be the reason
for not reproducing the very bottom of the line. The middle and bottom lines
show two hotter stars (spectral type O), only represented in {\it Martins} library. 
For this temperature range {\it Martins} library consider NLTE computations, and all 
Balmer profiles are very well reproduced.
 
\subsection{Summary}
\begin{figure*}
\resizebox{10cm}{!}{\includegraphics{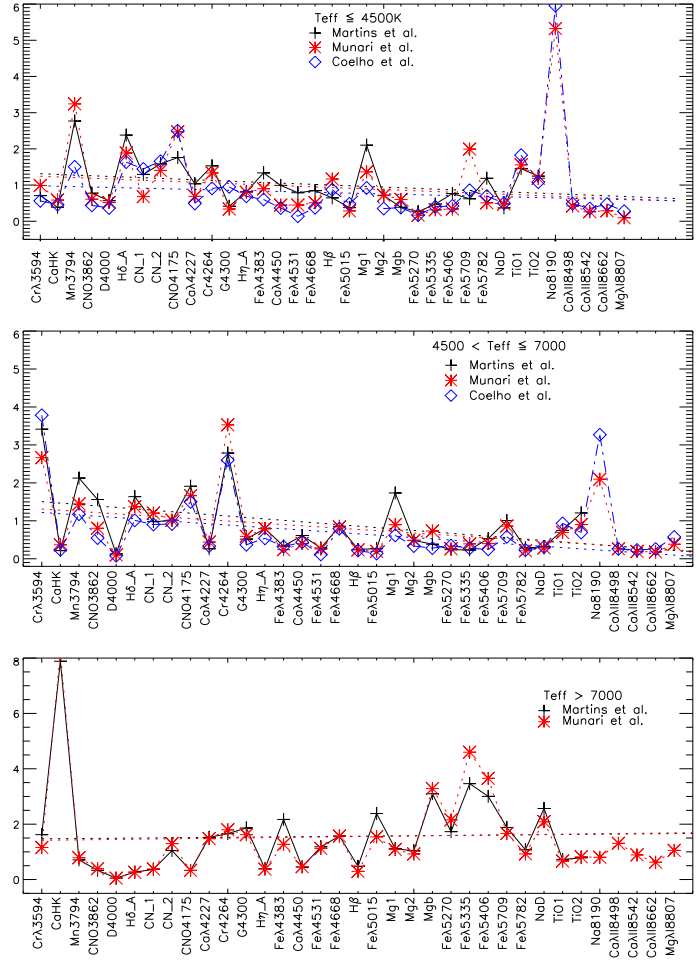}}
\caption{Average values of $adev$ for each index and each theoretical
library. The panels show three intervals of temperature, labelled in the plot. 
Each point is the
average $adev$ given by the comparison with the three empirical libraries.
Black crosses, red stars and blue diamonds represent the values for {\it Martins}, 
{\it Munari} and {\it Coelho} libraries respectively.
} 
\label{f:idx_adev}
\end{figure*}

The overall performance of the high resolution synthetic libraries is summarised in 
Figure \ref{f:idx_adev}. This figure shows the variation of $adev$ for each
theoretical library, split in the three {\Teff} intervals. We did not considered
observed stars that 
were significantly deviating from the other stars with similar {\Teff} and {\it log g}.
For each theoretical
library and each index, the $adev$ shown is the average of the $adev$ values obtained 
by the comparison to the three empirical libraries (the results for each 
of the empirical libraries are given in the Appendix).

The indices are shown on the $x$ axis, in order of
increasing wavelength. 
The dotted lines are linear fits of the $adev$ values for
each of the synthetic libraries 
(this fit does not take into account the near-IR indexes,
since the only empirical library that covers this region is Indo-US).
Although this figure cannot be seen as a precise
measure of the quality of the models, 
it can highlight interesting patterns.

First, all models are systematically deviating more in the 
blue part of the spectrum, where the blending of lines is considerably larger.
To improve the quality of the line list, specially in the blue region and further 
in the UV is the aim of the HST Treasury Program 9455 by Ruth Peterson
\citep[see e.g.][]{peterson+01,peterson+03}, 
and we confirm here that this is clearly the part of the spectrum that needs more work.

Second, {\it Coelho} library is the one that has the best average performance. 
This is likely
a consequence of their line list, which was calibrated along the years
in several high resolution stellar studies 
\citep[e.g.][]{erdelyi_barbuy89,castilho+99,melendez+03}. 
For stars hotter than 7000K {\it Martins} and {\it Munari} have similar results,
but again, these indices are very weak and provide almost no information on this hot
stars.
A visual comparison of the Balmer lines profiles shows, nevertheless, that above
{\Teff} $\sim$ 30000K, NLTE modelling is crucial.
 
The values of $adev$ are tabulated in the Tables in the
Appendix.

\section{Conclusions}

With this work we aimed at pointing
strengths and weaknesses of current theoretical stellar libraries, focusing 
on the observable values that are mostly used in stellar population models.

We divided our comparisons in two parts. 
In the first part, presented in \S3, we measured broad-band colours predicted by three of the most used 
model atmospheres grids currently available: \citet{ATLASODFNEW}, \citet{MARCS05}
and \citet{PHOENIX05}. We compared the model predictions with the recent empirical
colour-temperature relation by \citet{worthey_lee07}, for the stars that 
are representative of a young and a old simple stellar population.
The empirical relation is fairly well reproduced by the models for the colours V-I, 
V-R and J-H. All models are a little too blue in the B-V and H-K plane.
The biggest differences among the models, and also where they most deviate
from the empirical relation, is seen in the U-B colour. ATLAS9 is the model grid
that best represents the empirical relation (although a considerable improvement is still
required), but presents slightly higher residuals in the visual bands.
All colours of the cooler stars ({\Teff} $\apprle$ 4500K) also need improvements

The second part of our comparisons, presented in \S4, focus on the high 
resolution synthetic libraries that are aimed at spectroscopic studies.
We measured thirty five spectral indices defined in the literature on three
recent high resolution synthetic libraries: \citet{coelho+05}, \citet{martins+05}
and \citet{munari+05}. We compared the model indices with the observed measurements
given by three high quality empirical libraries by \citet{INDOUS}, \citet{MILES} and
\citet{ELODIE}.
Our first result is that it is not trivial to compare model and empirical
libraries because 
errors in the atmospheric parameters of the observed stars, and 
the particular abundance pattern of the solar neighbourhood 
might mislead into wrong conclusions about the accuracy of the models.
Overall, we found that models are systematically worse
in the same two regions that (not surprisingly) need improvement in the 
model atmospheres grids: the very cool stars and the blue region of the spectra. 
In the very low temperature regime
molecular and metallic lines dominate the spectra and models
are clearly struggling to reproduce them. A special care in the blue part
of the spectrum is needed.
It is worth highlighting the effect of different choices of atomic and molecular
line lists. The library by \citet{coelho+05} employs 
a line list that have been refined along the years in high resolution
stellar spectroscopic studies, and the effect is seen in its better average
performance, specially in the indices of iron peak elements.
The indices are hardly useful for evaluating hot stars. 
Only \citet{martins+05} and \citet{munari+05} libraries 
provide stars at the temperatures required for the 
modelling of young populations, and even then,
\citet{munari+05} is limited to {\it log g} $\geq$ 4.5, which
were not present in the empirical libraries.
A visual comparison of the Balmer lines profiles 
shows, as in previous studies, that NLTE modelling \citep[as in][]{martins+05} is required
for the good reproduction of the Balmer line profiles of very hot stars.
More tests are needed in this sense,
since these hot stars can have very different rotational velocities
which might affect the line profiles. Comparing models with 
slow rotation stars (which would not smooth the line profiles)
would be the best way to test how much models are indeed able
to reproduce these lines. 

As suggestions to the next generation of theoretical stellar libraries, we think that concerning the 
model atmospheres grids and flux distributions, 
a better reproduction of the blue flux (particularly important in the studies 
of young populations or any population at high redshift), and of the cool giants (that dominates 
the integrated spectra of old populations) are still required. It is unclear to the present
authors (none of us being a stellar atmosphere modeller) if this is to be achieved by more sophisticated
physics modelling (3D computations, NLTE, non-spherical effects, etc.), or improvements in the molecular, 
atomic and continuum opacities, or yet by adjusting model parameters (like mass, 
mixing-length parameter and micro-turbulent velocities) as a function of spectral type.
Concerning the high resolution libraries, we believe that significant improvement can still be made
by fine tuning the atomic and molecular line lists, through their calibration against high resolution
stellar spectra whose atmospheric parameters $-$ {\Teff}, {\it log g} and detailed abundance
ratios $-$ are known very accurately.

\vspace{1cm}

{\it Acknowledgements:}
PC is grateful to Guy Worthey for explaining many of the subtleties concerning the photometric
systems and the colour relations, 
and to Friedrich Kupka for the several and clarifying discussions about 
model atmospheres. LM thanks FAPESP through process 05/51101-5 for financial
support. We also thank the suggestions made by the anonymous referee
which helped  improve this manuscript.

\bibliographystyle{mn2e}

\bsp
\label{lastpage}

\end{document}